\documentclass[11pt,letterpaper]{article}

\pdfoutput=1  

\RequirePackage[OT1]{fontenc}
\RequirePackage{amsthm,amsmath}
\RequirePackage{natbib}
\RequirePackage[colorlinks,citecolor=blue,urlcolor=blue]{hyperref}

\usepackage{booktabs}
\usepackage{multirow}
\usepackage{siunitx}
\usepackage{palatino}
\usepackage[left=1in,top=1in,right=1in,bottom=1in]{geometry}
\usepackage{amsmath}
\usepackage{amssymb}
\usepackage{amsbsy}
\usepackage{amsthm}
\usepackage{epsfig}
\usepackage{clrscode}
\usepackage{setspace}
\usepackage{multirow}
\usepackage{url}
\usepackage{paralist}
\usepackage{wrapfig,boxedminipage}
\usepackage{graphicx}
\usepackage{subcaption}
\usepackage{float}
\usepackage{xcolor}
\usepackage{pgfplotstable}
\usepackage{hyperref}
\usepackage{enumitem} 
\usepackage{adjustbox}
\usepackage{booktabs}
\usepackage{csvsimple}
\usepackage{tikz}
\usepackage{dsfont}
\usepackage{wrapfig}
\usepackage{tikz}
\usetikzlibrary{shapes, calc, shapes, arrows, snakes}
\usetikzlibrary{positioning}
\newdimen\nodeDist
\tikzset{dist1/.style={path picture= {
    \begin{scope}[x=1pt,y=10pt]
      \draw plot[domain=-6:6] (\x,{\x/24});
    \end{scope}
    }
  }
}
\tikzset{dist2/.style={path picture= {
    \begin{scope}[x=1pt,y=10pt]
      \draw plot[domain=-6:6] (\x,{0.01*(\x-2)^2-0.4});
    \end{scope}
    }
  }
}
\tikzset{dist3/.style={path picture= {
    \begin{scope}[x=1pt,y=10pt]
      \draw plot[domain=-6:6] (\x,{0.85/(1 + exp(\x))-0.5});
    \end{scope}
    }
  }
}
\tikzstyle{f1}=[draw,circle,minimum size=25pt,inner sep=0pt,dist1]
\tikzstyle{f2}=[draw,circle,minimum size=25pt,inner sep=0pt,dist2]
\tikzstyle{f3}=[draw,circle,minimum size=25pt,inner sep=0pt,dist3]

\newcommand{\nc}{\newcommand}

\newcommand{\ldot}{\bullet}

\nc{\bX}{{\bf X}}
\nc{\lhat}[1][i]{\hat\lambda_{#1}^{-1(g)}}
\nc{\what}[1][j]{\hat\omega_{#1}^{-1(g)}}
\nc{\Li}{\hat\Lambda^{-1(g)}}
\nc{\Oi}{\hat\Omega^{-1(g)}}
\nc{\diag}[1]{\text{diag}\left(#1\right)}
\nc{\Siginv}{\Sigma^{-1}}
\nc{\Ominv}{\Omega^{-1}}
\nc{\bone}{{\bf 1}}

\newcommand{\x}{\mathbf{x}}

\newcommand{\beq}{\begin{equation}}
\newcommand{\eeq}{\end{equation}}

\newcommand{\ben}{\begin{enumerate}}
\newcommand{\een}{\end{enumerate}}

\providecommand{\keywords}[1]{\textbf{\textit{Keywords and phrases: }} #1}

\newcommand{\methN}{\text{BART with Targeted Smoothing}} 
\newcommand{\methA}{\text{tsBART}} 
\newcommand{\methAC}{\text{tsBART}} 

\begin{document}

\title{BART with Targeted Smoothing: An analysis of patient-specific stillbirth risk}

\author{Jennifer E.~Starling\footnote{Corresponding author: jstarling@utexas.edu}, Jared S. Murray, 
Carlos M.~Carvalho, \\ Radek Bukowski, and James G.~Scott}

\maketitle 

\begin{abstract}


This article introduces \methN{}, or \methA{}, a new Bayesian tree-based model for nonparametric regression.  The goal of tsBART is to introduce smoothness over a single target covariate $t$, while not necessarily requiring smoothness over other covariates $x$.  \methAC{} is based on the Bayesian Additive Regression Trees (BART) model, an ensemble of regression trees.  \methAC{} extends BART by parameterizing each tree's terminal nodes with smooth functions of $t$, rather than independent scalars.  Like BART, \methA{} captures complex nonlinear relationships and interactions among the predictors.  But unlike BART, tsBART guarantees that the response surface will be smooth in the target covariate.  This improves interpretability and helps regularize the estimate.

After introducing and benchmarking the \methA{} model, we apply it to our motivating example: pregnancy outcomes data from the National Center for Health Statistics.  Our aim is to provide patient-specific estimates of stillbirth risk across gestational age $(t)$, based on maternal and fetal risk factors $(x)$.  Obstetricians expect stillbirth risk to vary smoothly over gestational age, but not necessarily over other covariates, and \methA{} has been designed precisely to reflect this structural knowledge.  The results of our analysis show the clear superiority of the \methA{} model for quantifying stillbirth risk, thereby providing patients and doctors with better information for managing the risk of fetal mortality. All methods described here are implemented in the R package \textbf{tsbart}.

\end{abstract}



\section{Introduction}

An ongoing research challenge in obstetrics is to quantify the risk of stillbirth, defined as fetal death after 20 weeks of gestation.  Stillbirth is a major public-health problem, with 23,595 reported cases in the U.S.~in 2013 alone \citep{macdorman2015}.  Stillbirth is less well understood than other adverse pregnancy outcomes, and stillbirth rates have remained largely unchanged, even as many other serious adverse pregnancy outcomes (e.g.~neonatal death) have become rarer.  Providing better estimates of stillbirth risk as gestational age advances can yield important insights for obstetricians and patients.  If an obstetrician knew, for example, that a patient's stillbirth risk was likely to rise earlier in pregnancy than usual, or was likely to rise to higher than normal levels at later gestational ages, then proactive steps could be taken to manage that risk, especially in pregnancy at term.  Conservative steps might entail increased monitoring and more frequent prenatal clinic visits, while a more aggressive step might involve an elective Cesarean section or early induction of labor.

Statistically speaking, we can think of stillbirth risk as a regression function $h(t,x)$ representing the conditional probability\footnote{Or, in continuous time, the hazard rate.} of stillbirth at gestational age $t$, given that the fetus survived in utero until just before $t$, and given other characteristics $x$ of the maternal-fetal dyad.  Thus the fundamental biomedical problem we address in this paper is to provide better patient-specific estimates of $h(t,x)$.  This fills an important knowledge gap, since the current obstetrics literature does not provide an especially nuanced characterization of this function. In particular, the way that $h(t,x)$ depends upon maternal-fetal characteristics is not well understood. Structurally, obstetricians do expect that stillbirth risk evolves smoothly, without sudden jumps or discontinuities, as gestational age ($t$) advances; however, they do not have strong prior knowledge about how it should change with other maternal-fetal characteristics ($x$).

The central argument of our paper is that this situation calls for nonparametric regression with \textit{targeted smoothing} in gestational age $t$: that is, we require that $h(t, x)$ be smooth with respect to $t$ (the target covariate), but we remain agnostic about smoothness with respect to $x$.  This approach realizes two complementary advantages when quantifying stillbirth risk.  First, from a clinical perspective, targeted smoothing reflects prior knowledge, aids interpretability, and assists doctors in communicating stillbirth risks to patients as clearly as possible.  For example, smoothing helps prevent doctors and patients alike from over-interpreting the small jumps or wiggles in $h(t, x)$ that arise in a completely nonparametric estimate, but that are likely just noise.  Second, from a statistical perspective, targeted smoothing can reduce variance without inflating bias. 

To incorporate these benefits into our analysis of stillbirth risk, we propose a Bayesian approach called BART with Targeted Smoothing, or \methA{}, which is based on the highly successful Bayesian Additive Regression Trees (BART) model introduced by \cite{chipman2010}.  The original BART model is a Bayesian ensemble-of-trees approach to nonparametric regression.  It predicts a scalar response $y$ using a sum of many binary regression trees, where each tree is encouraged by a prior to be a ``weak learner''---that is, to have relatively few splits and to use only a small set of the available predictors.  \methN{} is similar in this regard, and we use the same prior over tree space proposed in the original BART paper.  Where \methA{} differs is in the prior used for the terminal nodes of each tree.  BART specifies a Gaussian prior for the scalar mean parameters in each terminal node.  \methAC{} replaces the Gaussian prior with a Gaussian process prior over univariate functions in the ``target'' covariate $t$, so that each terminal node is parameterized by a smooth function of $t$.

Thus to summarize our contributions:
\begin{enumerate}
\item We introduce the \methA{} model and demonstrate its advantages for problems where targeted smoothing is desirable.
\item We apply this method to data on birth records from the National Center for Health Statistics in order to produce accurate estimates for $h(t, x)$, providing clinicians with more granular knowledge of patient-specific stillbirth risk.
\end{enumerate}

It would certainly be possible to estimate stillbirth risk using existing techniques for modeling time-to-event data  \citep[see, e.g.][]{mandujano2013}.  Thus a major focus of our paper is to demonstrate that the specific features we had in mind when designing \methA{}---targeted smoothing in gestational age, while avoiding strong assumptions in other covariates---have some very real advantages for this kind of problem.  Available techniques either lack smoothness entirely (and thus tend to have smaller bias) or enforce smoothness globally (and thus tend to have smaller variance).  Each approach has its advantages, but \methA{} enjoys the best of both worlds for quantifying stillbirth risk: it easily handles complex interactions and non-linear effects, maintains computational tractability, and offers a full picture of posterior uncertainty, all while maintaining smoothness in $t$.

Moreover, while our motivating example involves estimating a smooth hazard function, the \methA{} model is much more general than this.  The same approach can work in any nonparametric regression problem where targeted smoothing is desired \textit{a priori}, regardless of whether the response is continuous, binary, or (as in our case) a time-to-event outcome.  Across a series of benchmarking examples, we show that our approach to targeted smoothing can lead to a favorable bias-variance tradeoff versus both classes of competing methods: those that make global smoothness assumptions, and those that make no smoothness assumptions.  Our simulation studies also bear out another considerable advantage: when the targeted smoothing assumption is valid, \methA{} tends to yield superior frequentist coverage versus plausible alternative methods.

The paper proceeds as follows.   Section 2 provides an overview of the stillbirth risk-curve modeling problem and dataset.  Section 3 details the \methA{} model and reviews the relevant literature. Section 4 presents the results of simulation studies showing the advantages of the method.  Section 5 then presents our core scientific contribution: an analysis of stillbirth risk using the \methA{} model.    Section 6 concludes with a brief discussion.  Further details, including on computational methods, are in the appendices.

All methods described in this paper are implemented in the R package \textbf{tsbart}.\footnote{https://github.com/jestarling/tsbart}

\section{Stillbirth Risk} \label{sec:data}  


\subsection{Background}

Stillbirth is a significant public health concern that affects tens of thousands of Americans each year. In the U.S. in 2013, a total of 23,595 stillbirths were reported \citep{macdorman2015}. The National Vital Statistics System notes that stillbirth has been significantly overlooked in public-health research and obstetrics guidance, and its mechanisms are not well understood.  Obstetricians do know that the risk of stillbirth typically (but not universally) cumulatively increases with time in utero.  But this risk must be balanced against the potential negative consequences of early delivery.  Preterm and early term births are associated with increased risk of neonatal mortality and morbidity, adverse neuro-developmental and cognitive outcomes, and increased healthcare costs \cite[e.g.][]{muraskas2008, kornhauser2010}.  Obstetricians can therefore benefit greatly from access to better estimates of stillbirth risk over gestational age, so that they can give clinical advice that minimizes the overall risk of adverse perinatal outcomes.  Conservative patient management might entail increased monitoring and more frequent prenatal clinic visits, while more aggressive steps include an early
delivery via either elective Cesarean section or early induction of labor.  From a statistical perspective, this means that accurate uncertainty quantification is vital, for helping doctors understand which cases have a less precisely estimated risk profile.

Previous research on adverse perinatal outcomes has focused more heavily on neonatal death than on stillbirth \cite[e.g.][]{bailit2010, clark2010, reddy2011}.  A more recent line of work attempts to refine these broad conclusions by seeking to model stillbirth risk based on a patient's individual risk factors.  In particular, \cite{mandujano2013} model hazard functions for stillbirth by stratifying patients into two broad categories: low risk versus high risk.  Here ``high risk'' is determined by presence or absence of at least one of several preexisting maternal conditions (e.g.~diabetes, chronic hypertension, and others). The model provides two stillbirth risk curves, one each for the high-risk and low-risk groups, for a U.S.~cohort.  This model does not meaningfully distinguish among the individual risk factors with potentially distinct etiologies, nor does it incorporate recent evidence that many other maternal and fetal characteristics---including maternal race, plurality, birth weight, and sex of the fetus---appear to correlate with stillbirth risk \citep{xu2013, macdorman2015}.  Finally, it fails to allow for the possibility of statistical interactions between risk factors.  Our targeted smoothing approach is specifically designed to address these shortcomings.


\subsection{Data description}

Our analysis uses anonymized birth data from the National Center for Health Statistics from the years 2004 and 2005 (Table \ref{tbl:cohort}).  Each medical record is associate with a single pregnancy.  It records contain the gestational age in weeks at which the pregnancy was delivered, based on calculation from the woman's last normal menstrual period, or a clinical estimate.  The outcome of each pregnancy is recorded as either a stillbirth or a live birth.  Each record also contains information about the maternal-fetal dyad, including maternal risk factors, such as diabetes, hypertension, and sociodemographic variables, and fetal characteristics, such as sex or estimated fetal weight.  

\begin{table}[t!]
    \caption{ \label{tbl:cohort}
    Cohort characteristics for our dataset on stillbirth.  The ``low risk'' and ``high-risk'' designations are not used formally in our model, but they are provided for the sake of comparison with \citet{mandujano2013}, Table 1.  Our cohort is similar in composition to the cohort analyzed there. Numbers in parentheses are percentages with respect to the given cohort.}
\centering
 \footnotesize
\begin{tabular}{lrrr}
                  & Full Cohort          & Low risk      & High risk  \\ 
   Characteristic &   (n=4,553,868) & (n=4,137,260) & (n=416,608) \\ 
    \toprule 
Maternal age (Yrs) &  &  &  \\ 
   \midrule
$<$20 & 452,060 (9.93) & 418,953 (10.13) & 33,107 (7.95) \\ 
  20-29 & 2,401,223 (52.73) & 2,204,168 (53.28) & 197,055 (47.30) \\ 
  30-39 & 1,585,226 (34.81) & 1,415,991 (34.23) & 169,235 (40.62) \\ 
  40-49 & 115,020 (2.53) & 97,855 (2.37) & 17,165 (4.12) \\ 
  50+ & 339 (0.01) & 293 (0.01) & 46 (0.01) \\ 
   \midrule
Maternal race and ethnicity &  &  &  \\ 
   \midrule
White, non-Hispanic & 2,757,816 (60.56) & 2,520,632 (60.93) & 237,184 (56.93) \\ 
  Black, non-Hispanic & 693,751 (15.23) & 619,761 (14.98) & 73,990 (17.76) \\ 
  Hispanic & 809,086 (17.77) & 736,908 (17.81) & 72,178 (17.33) \\ 
  Other & 293,215 (6.44) & 259,959 (6.28) & 33,256 (7.98) \\ 
   \midrule
Parity &  &  &  \\ 
   \midrule
Primiparous & 1,490,501 (32.73) & 1,370,443 (33.12) & 120,058 (28.82) \\ 
  Multiparous & 3,063,367 (67.27) & 2,766,817 (66.88) & 296,550 (71.18) \\ 
   \midrule
Maternal risk factors &  &  &  \\ 
   \midrule
Anemia & 115,663 (2.54) & 0 (0.00) & 115,663 (27.76) \\ 
  Cardiac disease & 20,937 (0.46) & 0 (0.00) & 20,937 (5.03) \\ 
  Lung disease & 63,063 (1.38) & 0 (0.00) & 63,063 (15.14) \\ 
  Diabetes mellitus & 159,765 (3.51) & 0 (0.00) & 159,765 (38.35) \\ 
  Hemoglobinopathy & 4,260 (0.09) & 0 (0.00) & 4,260 (1.02) \\ 
  Chronic hypertension & 43,935 (0.96) & 0 (0.00) & 43,935 (10.55) \\ 
  Renal disease & 14,210 (0.31) & 0 (0.00) & 14,210 (3.41) \\ 
  Rh isoimmunization & 31,317 (0.69) & 0 (0.00) & 31,317 (7.52) \\ 
   \midrule
Infant sex &  &  &  \\ 
   \midrule
Male & 2,330,557 (51.18) & 2,117,958 (51.19) & 212,599 (51.03) \\ 
  Female & 2,223,311 (48.82) & 2,019,302 (48.81) & 204,009 (48.97) \\ 
   \bottomrule
\end{tabular}
 \smallskip
\end{table}


The dataset consists of 8,371,461 pregnancies, with 7,940,495 live births, 100,072 stillbirths, and 330,894 cases where stillbirth outcome is missing.  We restrict our analysis to complete cases, with all maternal-fetal information and stillbirth response present.  Analysis is also limited to pregnancies delivered from 34 to 42 weeks inclusive, as is this the range where clinicians might plausibly recommend to deliver a baby based on elevated stillbirth risk, barring truly exceptional circumstances.  These restrictions yield 4,553,868 pregnancies for analysis, of which 7,175 are stillbirths, for an overall prevalence of 1.58 stillbirths per thousand pregnancies from 34 to 42 weeks' gestation. The prevalence in the high risk category was 2.85 stillbirths per thousand, while the prevalence in the low risk group was 1.45 per thousand.  Prevalence is comparable to the dataset analyzed by \citet{mandujano2013}, where overall prevalence was 1.45 births per thousand: 2.68 in the high risk group, and 1.34 in the low risk group.   A full table of summary statistics for our sample is shown in Table \ref{tbl:cohort}.  In practice, we work with a smaller case-control sample of this full data set.  This is described in Section \ref{sec:sbanalysis}; full details of the data pipeline are also available at \url{github.com/jestarling/tsbart-analysis}.

Maternal-fetal characteristics were selected for inclusion in our regression models based on clinical knowledge, availability of data, and previous research findings on risk factors for stillbirth \cite[e.g.][]{mandujano2013, muraskas2008, kornhauser2010}.  Maternal covariates include maternal age, primiparity, whether the labor was induced, ethnicity (White non-Hispanic, Black non-Hispanic, Hispanic, Other), aggregate pregnancy weight gain quantile, presence of diabetes mellitus, presence of chronic hypertension, and an indicator for the presence of any other risk factor.  Other risk factors include anemia, cardiac disease, lung disease, hemoglobinopathy, and Rh sensitization.  Consistent with the analysis of \citet{mandujano2013}, pregnancy-related complications, such as gestational diabetes, abruption, or preeclampsia, were not included as risk factors.  Fetal covariates include infant sex and birth weight quantile.

We did not exclude any variables on statistical grounds.  One of the benefits of the BART framework, which also applies to the \methA{} method, is that variable selection procedures are not generally required.  As discussed in Section \ref{sec:funbart}, the BART prior guides the model to choosing subsets of the most relevant covariates for inclusion in each tree.  

Birth weight cannot be observed directly by a doctor contemplating whether to delivery a pregnancy early due to elevated stillbirth risk.  However, birth weight quantile acts as a sensible proxy for the information doctors \textit{would} actually have at their disposal in a prenatal visit: fetal weight quantile in utero, which is estimated routinely using ultrasound and fetal growth charts.  Because fetal weight quantile at later gestational ages correlates very strongly with birthweight quantile, we do not expect that there is substantial error introduced by using birth weight quantile (which we have and a doctor wouldn't) as a proxy for fetal weight quantile in utero (which a doctor would have).

\section{\methN{}} \label{sec:funbart}  

We now introduce the \methA{} model, which later in Section \ref{sec:sbanalysis} we will use to analyze the stillbirth data just described.  Throughout the remaining sections, we let $t \in \mathcal{T}$ represent the target covariate, i.e.~the covariate in which the response surface is assumed to be smooth, which in our case is gestational age (discrete time, measured in weeks or days).  We let $x \in \mathcal{X}$ represent a vector of covariates other than $t$, which in our case are the characteristics of a particular maternal-fetal dyad. 

Because tsBART is a general approach for targeted smoothing in nonparametric regression, we first introduce the model in full generality.  We then explain how to adapt it more specifically for modeling the hazard function for stillbirth, $h(t,x)$, which represents the conditional probability of stillbirth at gestational age $t$, given that a fetus has survived in utero through gestational age $t-1$.  

~\\
\subsection{The BART model}

Before introducing  \methA{}, we briefly review the original BART framework.  BART (for Bayesian Additive Regression Trees) is a fully Bayesian ensemble-of-trees model \citep{chipman2010}.  BART models the mean response for a non-linear regression function as the sum of a large number of binary trees, each of which is constrained by the BART prior to be shallow (and therefore a weak learner).  The model is defined by a likelihood and prior, and inference is performed by sampling from the posterior.  Specifically, suppose that $y_i$ is a scalar response and $x_i$ is a vector of covariates.  The BART model assumes that
	\begin{align}
		y_i &= f(x_i) + \epsilon_i, \quad \epsilon_i \stackrel{iid}{\sim} \text{N}(0,\sigma^2) \\
		f(x_i) &= \sum_{j=1}^{m} g(x_i;T_j,M_j) \, .
	\end{align}
Here each $T_j$ is a binary tree that induces a step function in $x$ via a partition of the covariate space, while the $M_j = \left\{\mu_{1j},\hdots,\mu_{b_jj} \right\}$ are the $b_j$ terminal node values in tree $j$ (i.e.~the levels of the step function).  We can think of each $g$ as a basis function parameterized by the binary tree defined by $(T_j, M_j)$.

The BART prior consists of three elements.  The first component is the conjugate prior for the error variance, $\sigma^2 \sim   \nu \lambda / \chi^2_\nu$.  The second component is the specification of independent Gaussians $\mu_{hj} \stackrel{iid}{\sim} \text{N}(\mu_0,\tau^2)$ on the terminal node parameters $M_j = \left\{\mu_{1j},\hdots,\mu_{b_jj} \right\}$ of each tree.  The third component is the prior over tree space, composed of a set of probabilities governing three things: the choice of splitting covariate, the choice of splitting value for each covariate, and whether a node at a given depth is a terminal node.  We refer interested readers to \cite{chipman2010}, who recommend default hyperparameters that favor shallow trees, which both regularizes the estimate and encourages rapid mixing. 

BART has been successful in a variety of contexts including prediction and classification \citep{chipman2010, murray2017log, linero2017, linero2018, hernandez2018}, survival analysis \citep{sparapani2016}, and causal inference \citep{hill2011, hahn2017, logan2017, sivaganesan2017}.  

\subsection{The tsBART model}

\begin{figure}
	\includegraphics[width=0.9\textwidth]{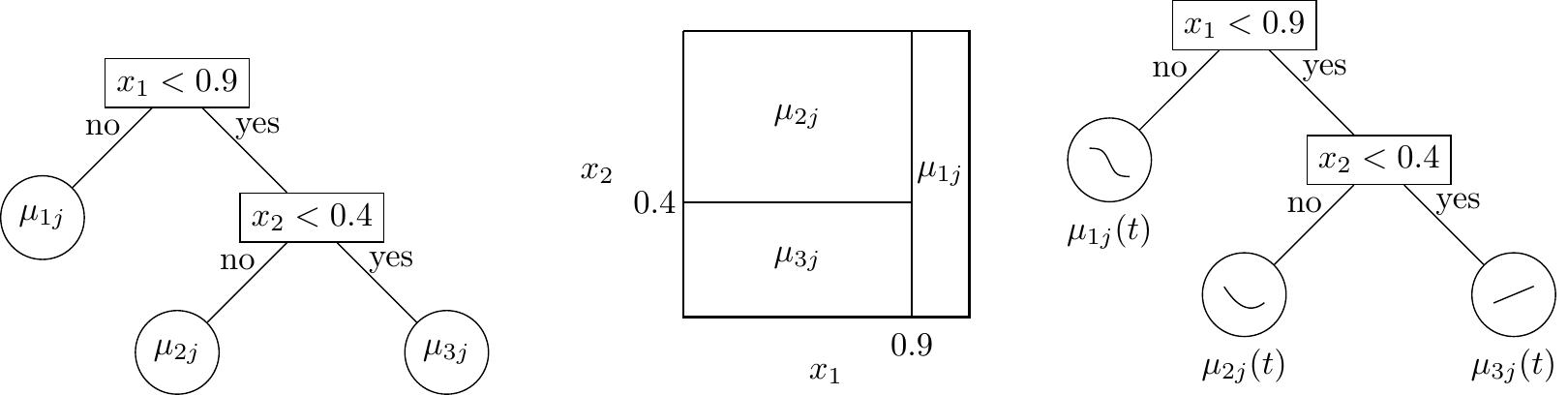}
	\caption[]{Illustrates the extension of BART to \methA{} for a single tree, where nodes are parameterized by smooth functions instead of scalar parameters.
				 (Left) A vanilla BART example binary tree $T_j$ where terminal nodes are labeled with the corresponding scalar parameters $\mu_{hj}$.
				 (Middle) The corresponding vanilla BART partition of the sample space and the step function $g(T_j, M_j)$.
				 (Right) Our \methN{} modification, where the $\mu_{hj}(t)$ parameters associated to terminal nodes are now functions of $t$.}
		\label{fig:funbart_tree}
	\end{figure}
	
Motivated by the success of BART models, we introduce \methA{}, an extension of BART for estimating regression functions that are smooth in a target covariate.  Consider a regression problem with scalar response $y_i = f(t_i, x_i) + e_i$, where the underlying mean function $f(t_i, x_i)$ depends both on $t$ (a scalar) and $x$ (a vector), and should be smooth in $t$.  To adapt BART for this setting, we replace the scalar node-level parameters $\mu_{hj}$ with univariate functions in $t$, $\mu_{hj}(t)$, and we assume that only $x$ variables (but not the target variable $t$) are used to define tree splits.  (See Figure~\ref{fig:funbart_tree}.) These univariate functions in $t$ can in principle be assigned any prior over function space; in the applications considered in this paper, we use Gaussian process priors.

More formally, we express the \methA{} model as follows.  Suppose that each observation $i$ in our data set consists of predictor variables $(t_i, x_i)$ together with outcome $y_i$ for $i = 1, \ldots, N$.  (Recall that $t_i$ is the target variable for smoothing, while $x_i$ is a vector of all other variables.) We now let
	\begin{align}
		y_{i} &= \alpha(t_i) + f(t_i,x_i) + \epsilon_{i}, \quad \epsilon_{i} \stackrel{iid}{\sim} \text{N}(0, \sigma^2) \label{eqn:funbart1} \\
		%
		f(t_i, x_i) &= \sum_{j=1}^{m} g(t_i, x_i; T_j, M_j) \nonumber \, .
	\end{align}
Here $T_j$ is a binary tree whose terminal nodes partition the ``non-target'' covariate space $\mathcal{X}$ into $b_j$ disjoint regions, just as in the original BART model.  But unlike BART, we parametrize the terminal nodes of the tree not by scalars, but by a collection of Gaussian processes in $t$: $M_j = \{\mu_{1j}(t), \ldots, \mu_{b_jj}(t) \}$, with each function $\mu_{hj}(t)$ associated with one terminal node.  The right panel of Figure \ref{fig:funbart_tree} illustrates an example with $b_j = 3$ terminal nodes.  The overall response is the sum of $m$ such trees, so that at any fixed design point $x = (x_1, \ldots, x_p)$, the response $f(t,x)$ is the sum of $m$ Gaussian processes.\footnote{This implies that $f(t, x)$ is a Gaussian process in $t$ for fixed $x$, but it is not a Gaussian process in $(t,x)$ jointly.}  We center the model at $\alpha(t)$, a baseline function of $t$, so that the trees parametrize deviations from the baseline that are associated with $x$. We estimate $\alpha(t)$ using the sample mean response for observations at each $t$.

We use the same prior over tree space as in the original BART paper.  To model the $\mu_{hj}(t)$'s in each terminal node, we use a zero-centered Gaussian process prior:
$$
\mu(t) \sim \text{GP}\left(0, C_\theta(t,t') \right) \, ,
$$
where $C_\theta(t, t')$ is the covariance function with hyperparameter $\theta$, which can be either chosen based on prior knowledge or tuned using the data.  (Zero-centering is appropriate here because we separate out the mean term $\alpha(t)$ in Equation \ref{eqn:funbart1}.)

In principle any covariance function can be used.  For all examples in this paper, we use the squared-exponential covariance function with variance parameter $\tau^2/m$ and length scale $l$.  That is,
\begin{equation}
\label{eqn:secovar}
C(t, t') = \frac{\tau^2}{m} \exp \left\{ - \frac{d(t, t')^2}{2l^2}   \right\} \, ,
\end{equation}
where $d(t, t')$ is the Euclidean distance between $t$ and $t'$.  Here $\tau^2$ determines the marginal variance of the $\mu_{hj}$'s, while $l$ governs their ``wiggliness.''  As in the original BART model, we scale the variance parameter $\tau^2$ inversely by the number of trees $m$.  Since the mean-response function $f(t,x)$ is the sum of $m$ trees, this implies that the marginal prior variance of $f(t,x)$ at any point $t$ is $\tau^2$.  We then assign $\tau$ a half-Cauchy prior as in \citet{gelman2006}, \citet{linero2017} and \citet{hahn2017}.

The \methA{} model also requires specifying $l$, the length scale of the Gaussian process prior, with larger $l$ corresponding to more wiggliness.  This length scale can be set using prior knowledge, but in Section \ref{ss:ecrosstune} we provide a method to tune it automatically over a grid of possible values.  As we also explain in Section \ref{ss:ecrosstune}, a reasonable default choice when using the squared exponential covariance function is $l = T/\pi$, where $T$ is the range of $t_{i}$ values in the data set. 

We make the simplifying assumption of an $i.i.d.$ error structure and complete the model specification by assigning $\sigma^2$ an inverse chi-square distribution $\sigma^2 \sim \nu \lambda / \chi^2_\nu$.  For full computational details, including the data augmentation, prior specification, and posterior full conditional distributions, see Appendices~\ref{appendix1} and \ref{appendix2}.

\subsection{Tuning the length scale $l$} \label{ss:ecrosstune}

We must select $l$, the length-scale parameter of the covariance matrix.  To do this, we represent $l$ using a formula by \citet{kratz2006} for the expected number of times a random function crosses its mean, $\text{E}\left[N_T\left(s \right) \right]$, on some interval $I = \left[0,t_\text{max} \right]$.  This formula gives us a closed-form solution for the length-scale parameter as a function of the expected number of times that $f(t,x)$ crosses zero.  Recall that if $f(t,x) = 0$, then the overall response at predictor $x$ is simply $\alpha(t)$, which we can think of as the baseline response over $t$.  The more times that $f(t,x)$ crosses zero, the more sharply the covariate-specific mean response deviates from the overall mean response.

To set $\text{E}\left[N_T\left(s \right) \right]$, let $r(s)$ be the correlation function between time $0$ and time $s$:
\begin{align*}
	r(s) = \frac{\text{E}\left[\left\{f(0,x) - \mu(0) \right\} \left\{f(s,x) - \mu(s) \right\} \right]}
	{\text{sd}\left(f(0,x) \right) \cdot \text{sd}\left(f(s,x) \right)}.
\end{align*}
Per Kratz,
\begin{align*}
	\text{E}\left[N_T\left(s \right) \right] = t_\text{max} \cdot \exp\left[-\frac{s^2}{2} \right] \left(\frac{\sqrt{r''(0)}}{\pi} \right)
\end{align*}
and we let $s=0$ in order to maximize $t_\text{max}$.  We use the squared exponential covariance kernel, so
\begin{align*}
	r(t) = \frac{\text{Cov}(f(0,x), f(t,x))}{\tau^2} = \exp\left[-\frac{t^2}{2l^2} \right].
\end{align*}
  Some algebra yields
\begin{align}
	l = \frac{t_\text{max}}{\pi \text{E}\left[N_T\left(0 \right) \right]} \, .
\end{align}

This opens up several options for choosing the length scale.  The first is by subjective choice.  This would entail eliciting a guess for $\kappa \equiv E[N_T(0)]$, the average number of times that $f(t,x)$ will cross zero over all values of the covariates---or equivalently, the average number of times that each response $\alpha(t) + f(t,x)$ will cross the overall mean response $\alpha(t)$.  This is a useful basis for elicitation, since the number of crossings is a sensible and intuitive measure for the wiggliness of our response as a function of $t$.

The second option is to choose a default value for $\kappa$.  If a default must be chosen, we recommend $\kappa=1$, or equivalently, $l = t_\text{max}/\pi$.  This encodes the belief that each response surface in $t$ will cross the overall mean response $\alpha(t)$ once, on average across all predictor values.  This allows for a substantial amount of heterogeneity in the mean responses over time, while still shrinking towards the overall mean.

A final option, which we use in our simulation studies and real-data examples, is to tune $\kappa = \text{E}\left[N_T\left(0 \right)\right]$ over a grid of candidate values.  This could be done using cross validation, as in the original BART paper, although we use the Watanabe–Akaike information criterion, or WAIC \citep{watanabe2013}.  WAIC is calculated as the log pointwise posterior predictive density plus a penalty for effective number of parameters, to avoid overfitting. It provides an estimate of generalization error without requiring that we split the data into multiple subsets; see Appendix~\ref{appendix3} for details.  In our simulation, we note that values of $\kappa \approx 1$ are frequently chosen by this data-driven approach, lending further credence to the choice of $\kappa = 1$ as a reasonable default.


\subsection{Adapting tsBART for binary and time-to-event outcomes}

\label{subsec:binary_hazard}

In their original paper, \citet{chipman2010} provide a probit version of the BART model for binary outcomes $Y \in \{0, 1\}$:
	\begin{align}
		p(Y=1 \mid x) &= \Phi\left(G(x) \right)  \\
		G(x) &= \sum_{j=1}^{m} g(x; T_j,M_j) \, ,
	\end{align}
where $\Phi(\cdot)$ is the standard normal CDF, and where $G$ is the standard BART model.  Inference proceeds via data augmentation, using the method of \citet{albert:chib:1993}.

Our \methA{} model can be extended in the same way.  Suppose that we observe a binary response $c_i$, together with target covariate $t_i$ and non-target covariates $x_i$.  The tsBART probit model introduces a latent Gaussian variable $z_i$, and then parametrizes $z_i$ using  tsBART, in a manner parallel to the original BART probit model:
	\begin{align}
	c_i &= \left\{
	\begin{array}{l l}
	1 & \mbox{if $z_i \geq 0$} \\
	0 & \mbox{if $z_i < 0$.} \\
	\end{array}
	\right. \\
		z_{i} &= \alpha(t_i) + f(t_i,x_i) + \epsilon_{i}, \quad \epsilon_{i} \stackrel{iid}{\sim}\text{N}(0, 1) \\
		%
		f(t_i, x_i) &= \sum_{j=1}^{m} g(t_i, x_i; T_j, M_j) \, .
	\end{align}
Here $\alpha(t)$ and $f(t, x)$ are defined exactly as in Equation \ref{eqn:funbart1}, and each $g_j$ is assigned the same prior outlined the previous subsection.  Marginalizing over $z_i$ yields the desired probability under the probit model, $P(c_i = 1 \mid x_i, t_i) = \Phi \{  \alpha(t_i) + f(t_i, x_i) \}$.

Crucially for our application, it is also straightforward to extend tsBART-probit to discrete right-censored time-to-event outcomes, as noted by \citet{sparapani2016} in the context of the original BART-probit model.  Suppose that $t_i \in \mathcal{T}$ is a discrete time-to-event outcome, and that $c_i$ is a censoring indicator: $c_i = 1$ means that an event occurred at time $t_i$, while $c_i = 0$ means that observation $i$ was right-censored at time $t_i$.  In our stillbirth risk-modeling problem, $c_i = 1$ corresponds to a stillbirth at gestational age $t_i$, while $c_i = 0$ corresponds to a live birth at $t_i$ (which is right-censoring with respect to the stillbirth event).  The object of interest is the set of conditional probabilities $\mathbf{p} = \{p_{it}\}$, where $p_{it}$ the conditional probability of an event at time $t$ for observation $i$, given than no event has happened through time $t-1$.  These conditional probabilities define the discrete-time hazard function $h(t, x)$.  For ease of exposition, we assume here that the possible event times are $\mathcal{T} = \{1, \ldots, T\}$, but this is not a requirement.

To accommodate this data structure, we use the following standard factorization of the likelihood for a discrete-time hazard model.  We introduce binary auxiliary variables $\{\tilde{c}_{is}: s = 1, \ldots t_i\}$ for each observation $i = 1, \ldots, N$, where
$$
\tilde{c}_{is} =  \left\{
	\begin{array}{l l}
	1 & \mbox{if $c_i = 1$ and $s = t_i$} \\
	0 & \mbox{otherwise.} \\
	\end{array}
	\right. \\
$$
The likelihood for the hazard function is now
$$
L(\mathbf{p}) = \prod_{i=1}^N \prod_{s=1}^{t_i} p_{is}^{\tilde{c}_{is}} \ (1 - p_{is})^{1- \tilde{c}_{is}} \, .
$$
We note, as do \citet{sparapani2016}, that the product form of this likelihood does not come from the assumption that the binary $\tilde{c}_{is}$ events are independent, but rather from the definition of each $p_{is}$ as a conditional probability.

We can now use the same latent-variable trick from \citet{albert:chib:1993} to construct the tsBART-probit model for $\mathbf{p}$, as follows:
\begin{align}
	\tilde{c}_{is} &\sim \left\{
	\begin{array}{l l}
	1 & \mbox{if $z_{is} \geq 0$} \\
	0 & \mbox{if $z_{is} < 0$.} \\
	\end{array}
	\right. \\
z_{is} &= \alpha(s) + f(s, x_i) + \epsilon_{is}, \quad \epsilon_{is} \stackrel{iid}{\sim} \text{N}(0, 1) \\
f(s, x_i) &= \sum_{j=1}^{m} g(s, x_i; T_j, M_j) \, ,
\end{align}
where $\alpha$ and the $g_j$'s are parametrized just as in the tsBART model described previously, treating time as the target covariate for smoothing. 

\subsection{Connection with existing work} \label{ss:exwork}

Our paper sits in a long line of other research on extensions to the Bayesian tree-modeling framework.  Two papers in particular are especially close in spirit to ours.  The first is \cite{sparapani2016}, who introduce a model for nonparametric survival analysis using BART.  Their model incorporates dependence on $t$ by simply adding time as an ordinary covariate to a BART-probit for the discrete-time hazard function.  This does not impose any continuity or smoothness constraints on $f(t,x)$.  In contrast, our approach smooths the hazard function over time, while still retaining the  benefits of BART.  The second paper is the treed Gaussian process (TGP) model of \citet{gramacy2008}.  Their model uses a single deep tree with a Gaussian process in each terminal node; our model, in contrast, is a sum of many trees.  Our work therefore generalizes that of \citet{gramacy2008} in the same way that the BART model generalizes the single-tree Bayesian CART model of \citet{chipman1998}.  

Smooth or partially smooth extensions of Bayesian tree models have also been proposed previously by \citet{linero2017}, who smooth a regression tree ensemble by randomizing the decision rules at internal nodes of the tree. This model induces smoothness over \textit{all} covariates by effectively replacing the step function induced by the binary trees with sigmoids.  In contrast, our approach smooths over just one target covariate, while avoiding the high computational cost associated with the method of  \citet{linero2017}.

\section{Simulations} \label{sec:simstudy} 

We conduct two simulation studies to compare \methA{} to existing methods.  These simulations are designed to evaluate tsBART along several dimensions---out-of-sample predictive performance, credible interval coverage, and interpretability---in settings with varying degrees of complexity in covariate interactions.

Given the importance of uncertainty quantification for modeling stillbirth risk, we do not benchmark against pure machine-learning methods that do not readily produce valid confidence or credible intervals.  This excludes neural networks, boosting, CART, and many other ensemble methods. We do, however, benchmark against BART, which has been shown to enjoy comparable or superior predictive performance to all these pure machine-learning methods across a range of scenarios \citep[see, e.g.][who run these comparisons across 42 benchmark data sets]{chipman2010}.  Thus very little is lost by excluding methods that perform comparably to BART in terms of pure prediction, but that cannot produce confidence/credible sets for those predictions. 

One plausible benchmark might be Random Forests, for which recent research \citep{wager2014} has addressed the problem of accurate uncertainty quantification.  However, we choose not to include Random Forest in the simulation benchmarks for two reasons.  First, \citet{chipman2010} performed extensive benchmarking of ordinary BART versus Random Forests, and they make a persuasive argument that if the computational resources are available for BART, it tends to perform a bit better, on average.  Additionally, we did investigate the performance of Random Forests on the stillbirth dataset that we analyze in Section 5.  We found that the stillbirth risk curve estimates provided by Random Forest had many of the same interpretational problems posed by BART---namely, by not imposing adequate smoothness over time, it limits the interpretability for clinicians, encouraging them to over-interpret small wiggles in the fit.  This analysis is included in Appendix~\ref{appendix5}.


~\\
\subsection{Simulation 1 - Direct Comparison with BART}

We first conduct a simulation study comparing \methA{} to the ordinary BART model.  The initial focus on BART is intended to isolate a key feature of our approach: smoothing in $t$, versus simply including $t$ as another predictor available in the model.  BART is also the most relevant practical comparison for our application, since \citet{sparapani2016} have already shown that ordinary BART-probit has cutting-edge performance for discrete-time survival modeling, versus a wide range of competing methods, including many more traditional time-to-event models.

We simulated datasets across three scenarios of modest dimension in the non-target variables $x$: one with four covariates, one with eight covariates, and another with twenty covariates.  For all scenarios, we used eight discrete time points ($\mathcal{T} = \{1, \ldots, 8\}$) for the target covariate.  This mimics the stillbirth data, where information on gestational age is used at a weekly resolution between 34 and 42 weeks.  It also reflects many other obstetrics, public health, and biomedical applications where data is observed at discrete intervals.  We generated each pair of covariates $(x_{ij}, x_{i,j+1})$ for odd $j$ from a bivariate Gaussian with moderate correlation and unit variances.  For each case, we simulated data sets with sample sizes $n \in \left\{100, 500, 1000, 2500 \right\}$, for a total of twelve combinations of sample size ($n$) and dimension of the non-targer covariate ($p$).  For each of these twelve combinations, we simulated 100 datasets.

We focus on a ground truth in which the mean response evolves smoothly in $t$, and we seek to answer two key questions: 1) can tsBART adapt to the correct degree of smoothness, and 2) if so, how large are the gains versus an otherwise very similar model that makes no smoothness assumptions?  In the $p=4$ case, we let
\begin{align*}
	f(t,x) = g(x_1,x_2) \cdot \cos\left(t + 2\pi h(x_3,x_4) \right)
\end{align*}
so that the covariates $x$ modify both amplitude and phase shift.  We let $g$ and $h$ be simple functions of the covariate pairs; here we sum each pair of covariates.

In the $p=8$ and $p=20$ cases, we continue in a similar fashion, alternating sines and cosines, so that
\begin{align*}
	f(t,x) = g(x_1,x_2) \cdot \cos\left(t + 2\pi h(x_3,x_4) \right) +
	g(x_5,x_6) \cdot \sin\left(t + 2\pi h(x_7,x_8) \right) + \hdots
\end{align*}
where this pattern continues.  We again let $g$ and $h$ be sums of each pair of covariates.  We generate responses $y(t,x) = f(t,x) + \epsilon$ where $\epsilon \stackrel{iid}{\sim}\text{N}(0,1)$.

We compare BART and \methAC{} using $m=200$ trees and 10,000 MCMC draws, with a burn-in of 1000 draws.  We compare performance by calculating the log-loss at each iteration of the algorithm, both in-sample and for a held-out sample, taking the mean log-loss across all MCMC iterations. Log-losses are scaled by sample size.  We tune the length scale $l$ using the method described in Appendix~\ref{appendix3}.  We compare models using log-loss since our goal is not to classify patients by whether they will experience stillbirth, but to provide well-calibrated probabilities of stillbirth to clinicians.  Log-loss is a proper scoring rule which measures how effectively each method calibrates its probability estimates.

	\begin{table}
	\caption{In-sample and out-of-sample log-loss (scaled by sample size), averaged across 100 replicates, comparing BART with tsBART.  \methAC{} consistently outperforms ordinary BART in the out-of-sample log-loss.  \methAC{} has the most significant gains in scenarios with small sample sizes or more predictors, as evidenced by p-values from a paired Wilcoxon test comparing out-of-sample results.}
		\centering
		\begin{tabular}{llccccc}
			\toprule
			  & &    \multicolumn{2}{c}{In-sample}      &   \multicolumn{3}{c}{Out-of-sample} \\
			p &		n &		BART &				tsBART  &							BART &					tsBART	&			P-value		   \\
			\midrule
			4 &		100 &	-1.61 &				\textbf{-1.49} &					-1.97 &					\textbf{-1.92} &	$<$0.001		\\ 
			4 &		500 &	-1.53 &				\textbf{-1.47} &					-1.80 &					\textbf{-1.74} &	$<$0.001		\\ 
			4 &		1000 &	-1.47 &				\textbf{-1.46} &					-1.76 &					\textbf{-1.72} &	0.001		\\ 
			4 &		2500 &	\textbf{-1.43} &	-1.44 &								-1.68 &					\textbf{-1.67} &	0.367		\\ 
			8 &		100 &	-1.74 &				\textbf{-1.66} &					-2.18 &					\textbf{-2.07} &	$<$0.001		\\ 
			8 &		500 &	-1.66 &				\textbf{-1.63} &					-2.02 &					\textbf{-1.92} &	$<$0.001		\\ 
			8 &		1000 &	\textbf{-1.55} &				-1.58 &					-1.95 &					\textbf{-1.91} &	0.002		\\ 
			8 &		2500 &	\textbf{-1.48} &				-1.53 &					-1.88 &					\textbf{-1.87} &	0.742		\\ 
			20 &	100 &	-2.04 &				\textbf{-2.02} &					-2.59 &					\textbf{-2.31} &	$<$0.001		\\ 
			20 &	500 &	\textbf{-1.94} &				-1.99 &					-2.41 &					\textbf{-2.28} &	$<$0.001		\\ 
			20 &	1000 &	\textbf{-1.81} &				-1.94 &					-2.31 &					\textbf{-2.27} &	$<$0.001		\\ 
			20 &	2500 &	\textbf{-1.66} &				-1.84 &					\textbf{-2.27} &		-2.32 &				0.023		\\
			\bottomrule
		\end{tabular}
					\label{tbl:simstudy}
	\end{table}
	
	\begin{figure}
			\centering
				\includegraphics[width=0.9\textwidth]{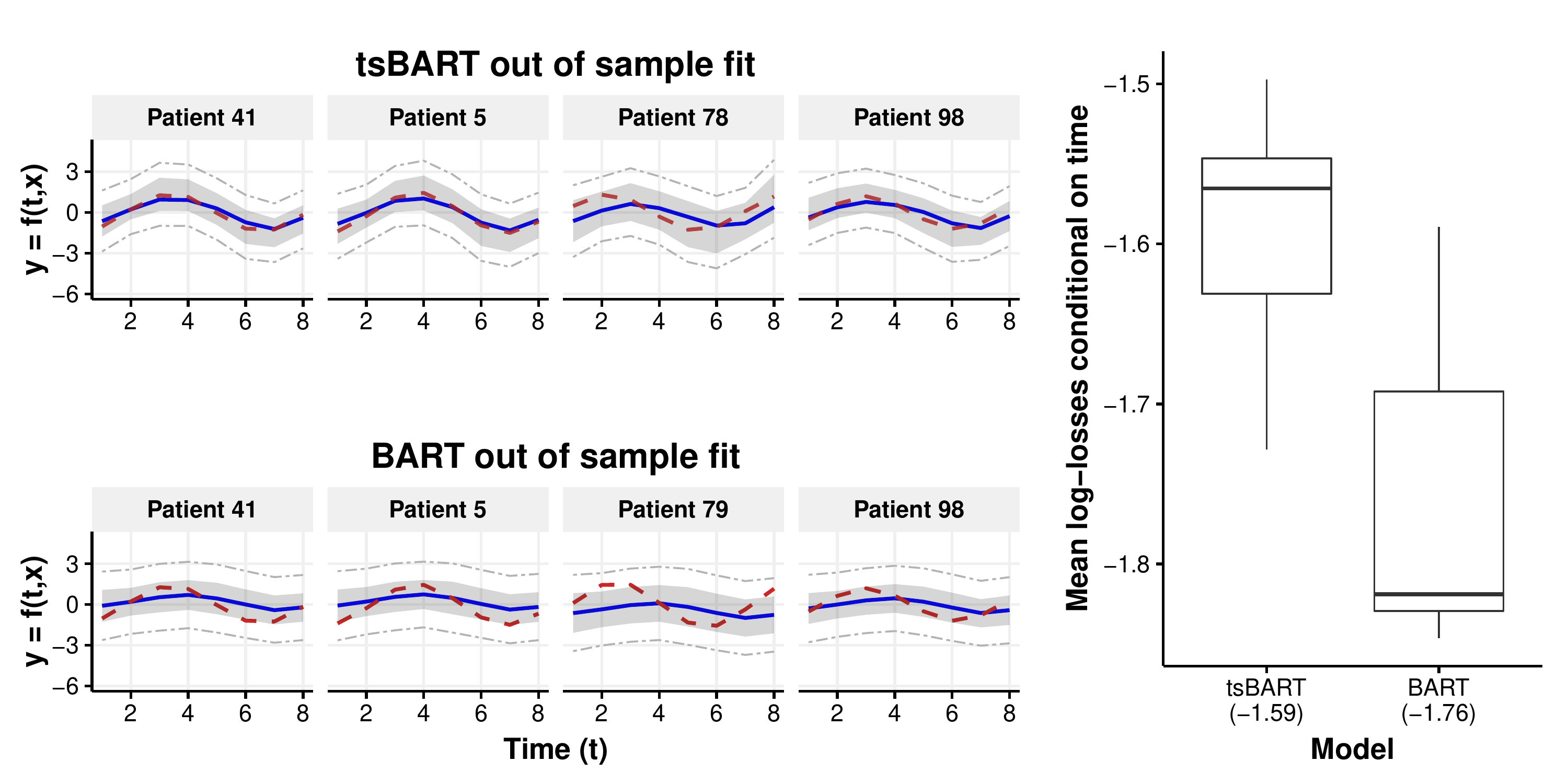}
			\caption{Compares a single model fit for \methA{} and BART, for one scenario (n=500, p=4), to illustrate the difference in fit when the ground truth is smooth in a single target covariate $t$. (Left) The bold dashed line represents the true function value, while the solid line and shading give posterior means and 95\% credible intervals.  Lighter dashed lines give the prediction intervals.  (Right) \methAC{} outperforms BART across $t$ and on average.  The boxplot gives the distribution of average log-loss at each $t$ for both methods, with overall average log-loss in parenthesis on the x-axis label.}
			\label{fig6}
		\end{figure}
	
\methAC{} consistently outperforms ordinary BART (Table \ref{tbl:simstudy}) in the out of sample log-loss.  \methAC{} has the most significant gains in scenarios with small sample sizes or more predictors.  Figure \ref{fig6} illustrates the out of sample fits and log-loss in a single scenario, where $n=500$ and $p=4$; \methA{} tends to smooth out the long-range periodicities in $f(t,x)$ much less than ordinary BART.  

\subsection{Simulation 2 - Comparison with BART and Splines}

We next compare \methA{} to four existing models in a simulation study designed to mimic the basic properties of the hazard functions we expect to see in our stillbirth data.   We generate hazard functions and corresponding survival data for three scenarios, where covariates determine shape of the hazard function with increasing degrees of interaction complexity.  We compare the following methods.
\begin{enumerate}
\item tsBART:  The \methN{} method with smoothing parameter $\kappa$ tuned as described in Appendix~\ref{appendix3}.
\item tsBART (default): The \methN{} method with our suggested $\kappa=1$ default smoothing parameter.
\item BART: an ordinary BART-probit model, which also sets hyperparameters ($\nu,\lambda$) as recommended in \cite{chipman2010}, and includes time $t$ as a covariate (BART)
\item Splines 1: a logistic regression model using cubic B-splines with seven degrees of freedom, with main effects for all covariates included in $x_i$.  Use of the spline basis induces targeted smoothness in $t$ by ensuring that, for fixed $x$, $f(t,x)$ is piecewise polynomial with continuous first and second derivatives.
\item Splines 2: another logistic regression model using cubic B-splines and seven degrees of freedom, with the addition of interactions between each basis element in $t$ and each covariate in $x_i$.
\item P-Splines: a penalized spline model including the same covariates and a penalized spline basis with 9 spline basis elements and a second-order smoothing penalty (P-splines) \citep{eilers1996}.  (The maximum possible number of basis elements is determined by the fact that there are only 9 distinct values for gestational age, 34--42 weeks.)  By allowing for all possible basis elements to enter the model while penalizing deviations from smoothness, penalized splines provide flexibility while still regularizing the stillbirth risk curve estimates.
\end{enumerate}
We evaluate the performance of \methA{} for three scenarios, representing increasing degrees of difficulty in how $x$ parametrizes the hazard function.

We simulate data as follows.  Let $t$ be a grid of times on the unit interval, spaced in increments of 0.1.  Generate $n=1000$ ten-dimensional covariates $x_i = \left\{x_{i1}, \hdots, x_{i10} \right\}$ where $x_{ij} \stackrel{iid}{\sim} \text{U}(0,1)$.  The first five covariates in each $x_i$ impact the response; the rest are noise.  In each case, we define the hazard function as the weighted combination of two ``template'' hazard functions $f_1(t)$ and $f_2(t)$, where weights $w(x_i)$ depend on covariates $x_i$:
\begin{align*}
	h(t, x) = 0.25x_{i5} + w(x_{i1}, \ldots, x_{i4})f_2(t) + \left[1- w(x_{i1}, \ldots, x_{i4}) \right] f_1(t) \, .
\end{align*}

\begin{figure}
		\centering
			\includegraphics[width=.8\textwidth]{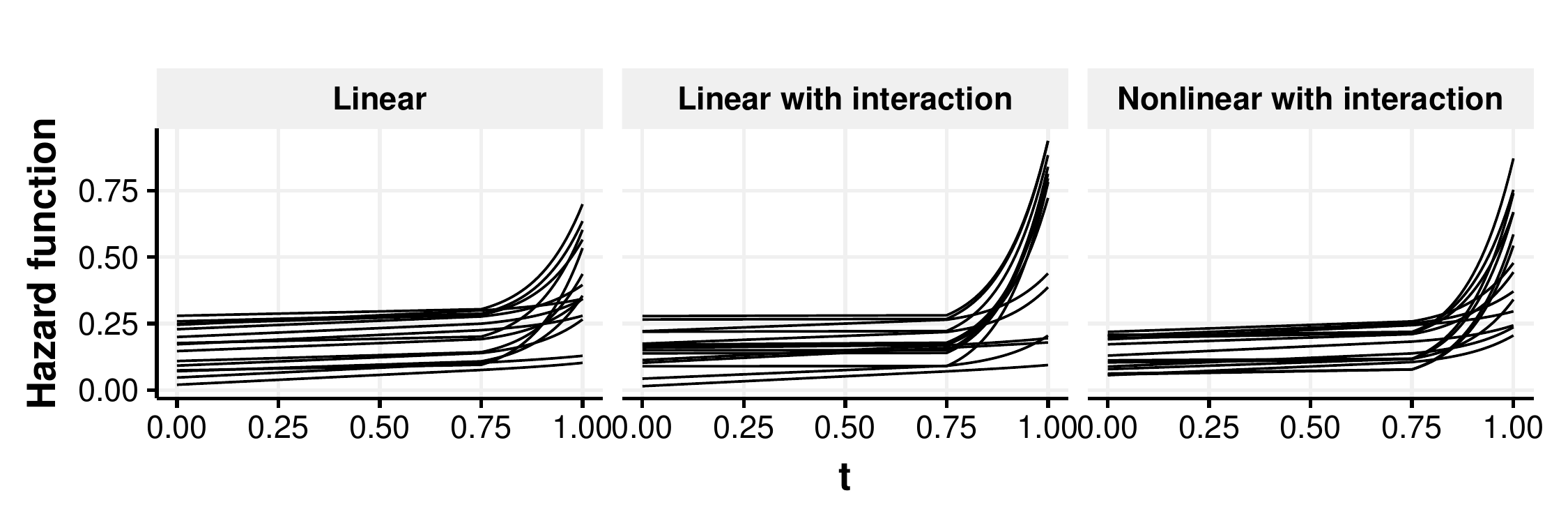}
		\caption{Illustrates the different ground truth hazards used in the three simulation scenarios. The same basic shapes of hazard functions are present in all three scenarios; the difference is in how covariates $x$ influence which shape arises.  There is a mixture of gently-rising hazards and hockey stick hazards; linearity determines the straightness of the rise, and presence of an interaction increases strength of the hockey stick.}
		\label{fig:sim}
	\end{figure}

The differences between the three scenarios are in how the weight depends on the covariates: linearly, linearly with interactions, or nonlinearly with interactions.  Figure \ref{fig:sim} illustrates resulting hazard functions for each scenario.  There are four general hazard function shapes, dictated by high versus low baseline risk, and with or without a sharp increase in hazard beginning at $t = 0.75$.  Appendix~\ref{appendix4} provides further detail, and code is available at \url{https://github.com/jestarling/tsbart-analysis}.  

\begin{singlespacing}
\begin{table}
	  	\caption{Average coverage rates (nominal is 95\%) and mean squared error across 500 simulated datasets for each weighting scenario and model combination. For tsBART and BART, coverage is for posterior credible intervals and mean squared error uses the posterior mean.  For the spline-based methods, coverage is for prediction intervals. \methAC{} has better coverage, even with the default smoothing parameter, and MSE for all methods is small and comparable.}
\centering
\begin{tabular}{llrr}
	\hline
 Weighting scenario & Method & Coverage & MSE \\ 
  \hline
Linear & tsBART (tuned) & 0.9310 & 0.0014 \\ 
    & tsBART (default) & 0.9092 & 0.0017 \\ 
    & BART & 0.7642 & 0.0011 \\ 
    & Splines 1 (Linear) & 0.7925 & 0.0007 \\ 
    & Splines 2 (Interaction) & 0.7788 & 0.0014 \\ 
    & P-Splines & 0.7720 & 0.0007 \\ 
   \hline
Linear (with interaction) & tsBART (tuned)& 0.9571 & 0.0019 \\ 
    & tsBART (default) & 0.9443 & 0.0022 \\ 
    & BART & 0.7907 & 0.0022 \\ 
    & Splines 1 (Linear) & 0.8874 & 0.0039 \\ 
    & Splines 2 (Interaction) & 0.7213 & 0.0391 \\ 
    & P-Splines & 0.8718 & 0.0036 \\ 
   \hline
Nonlinear (with interaction) & tsBART (tuned)& 0.9539 & 0.0013 \\ 
    & tsBART (default) & 0.9354 & 0.0016 \\ 
    & BART & 0.7408 & 0.0012 \\ 
    & Splines 1 (Linear) & 0.8918 & 0.0006 \\ 
    & Splines 2 (Interaction) & 0.8392 & 0.0013 \\ 
    & P-Splines & 0.8747 & 0.0006 \\ 
   \hline
\end{tabular}
  	\label{tab:simtab}
\end{table}
\end{singlespacing}

For each of the three scenarios, we simulate 500 datasets to compare point-wise coverage of \methA{} compared to the methods detailed in Section \ref{sec:sbanalysis}.  The mean-squared error of the estimates are small and comparable across all methods.  Most striking, however, is that \methAC{} gives far better coverage than other methods, both with the smoothing parameter tuned and set to the default value of 1 (Table \ref{tab:simtab}).  No other method consistently produces credible/confidence sets that are close to the nominal value of 95\%.  We conclude that tsBART is capable of matching or exceeding other methods in terms of mean-squared error, while producing error bars that are statistically trustworthy and scientifically sensible.

\section{Results for Modeling Stillbirth Risk} \label{sec:sbanalysis} 
	
We now turn to our motivating application, by applying the \methA{} method to estimate patient-specific stillbirth risk, using the data described in Section \ref{sec:data}.  To model the discrete-time hazard function for stillbirth, $h(t,x)$, we use the extension of tsBART-probit formulation described in Section \ref{subsec:binary_hazard}.  Our target covariate for smoothing is gestational age in weeks: $t_i \in \left\{34,35,\hdots,42 \right\}$.  We let $y_i$ be an indicator of whether stillbirth has occurred for each pregnancy, and $x_i$ be the vector of maternal-fetal covariates for each patient, including maternal age, primiparity, ethnicity, infant sex, presence of diabetes mellitus, presence of chronic hypertension, presence of other risk factors, whether the pregnancy was induced, and birth weight and weight gain quantiles.
 
We first focus on the question of whether tsBART does, indeed, yield better-calibrated risk estimates over existing methods for our data set.  For the purpose of evaluating all models while maintaining computational tractability, we created five balanced case-control samples of $n=1,000$ pregnancies each.  (Since stillbirth is a rare event, using a balanced case-control sample also more clearly highlights differences among methods.)  We then split each balanced case-control sample into training and testing sets. We used the training set to fit \methA{}, in addition to each of the four models discussed in Section \ref{sec:simstudy}: vanilla BART, the two B-spline models, and P-splines.  We tune the length-scale parameter of \methA{} using the method described in Appendix~\ref{appendix3}., and we set tree-prior hyperparameters ($\nu,\lambda$) as recommended in \cite{chipman2010}.  We then used the fitted model to predict the hazard functions for all held-out points, and we computed held-out log-losses. We repeat this process over five balanced case-control data sets and average the results (Table \ref{tbl:binll}).

\begin{table}
	\caption{Overall out-of-sample log-loss for each method, averaged over five evenly balanced case-control samples.  \methAC{} outperforms other methods, with the tuned smoothness parameter only slightly outperforming the default.}
\centering
\begin{tabular}{lc}
  \toprule
	Method & Log-loss \\ 
  \hline
   tsBART (tuned) & \textbf{-1.711} \\
   tsBART (default) & -1.713 \\
   BART & -1.810 \\
   Splines 1 & -1.725 \\
   Splines 2 & -1.919 \\
   P-splines & -1.724 \\
   \bottomrule
\end{tabular}
\label{tbl:binll}
\end{table}

\begin{figure}
		\centering
			\includegraphics[width=0.4\textwidth]{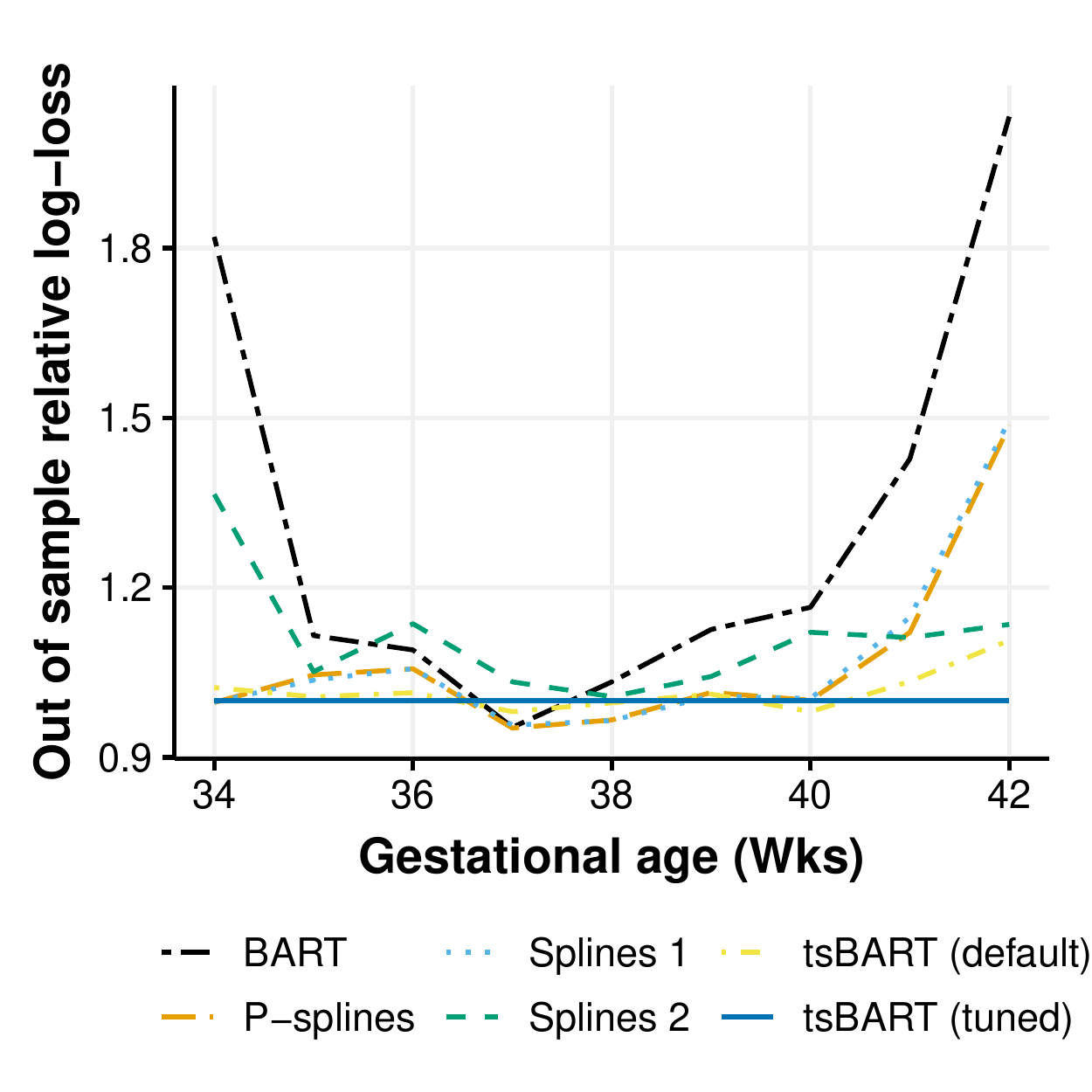}
		\caption{Illustrates performance of each method relative to \methA{} with tuned smoothing parameter $\kappa$.  Shows weekly out-of-sample log-loss for each method, averaged over five evenly balanced case-control samples. \methAC{}'s gains are especially apparent in higher and lower gestational ages; other methods have small gains in the 37 to 39 week range, at the expense of inflation at extreme gestational ages where sample sizes are small.}
		\label{fig:fig3}
	\end{figure}

\methAC{} outperforms other methods, with the tuned smoothness parameter setting only slightly outperforming the default (untuned) setting.  To provide some intuition for these results, Figure \ref{fig:fig3} also shows relative out-of-sample log-losses of all methods as a function of gestational age, with tuned \methA{} normalized to 1.  The figure shows that \methA{}'s gains are especially apparent at higher and lower gestational ages, where fewer observations are available.  Most methods are comparable at gestational ages across the middle of the available range (37-39 weeks).

\begin{figure}[t!]
		\centering
			\includegraphics[width=0.75\textwidth]{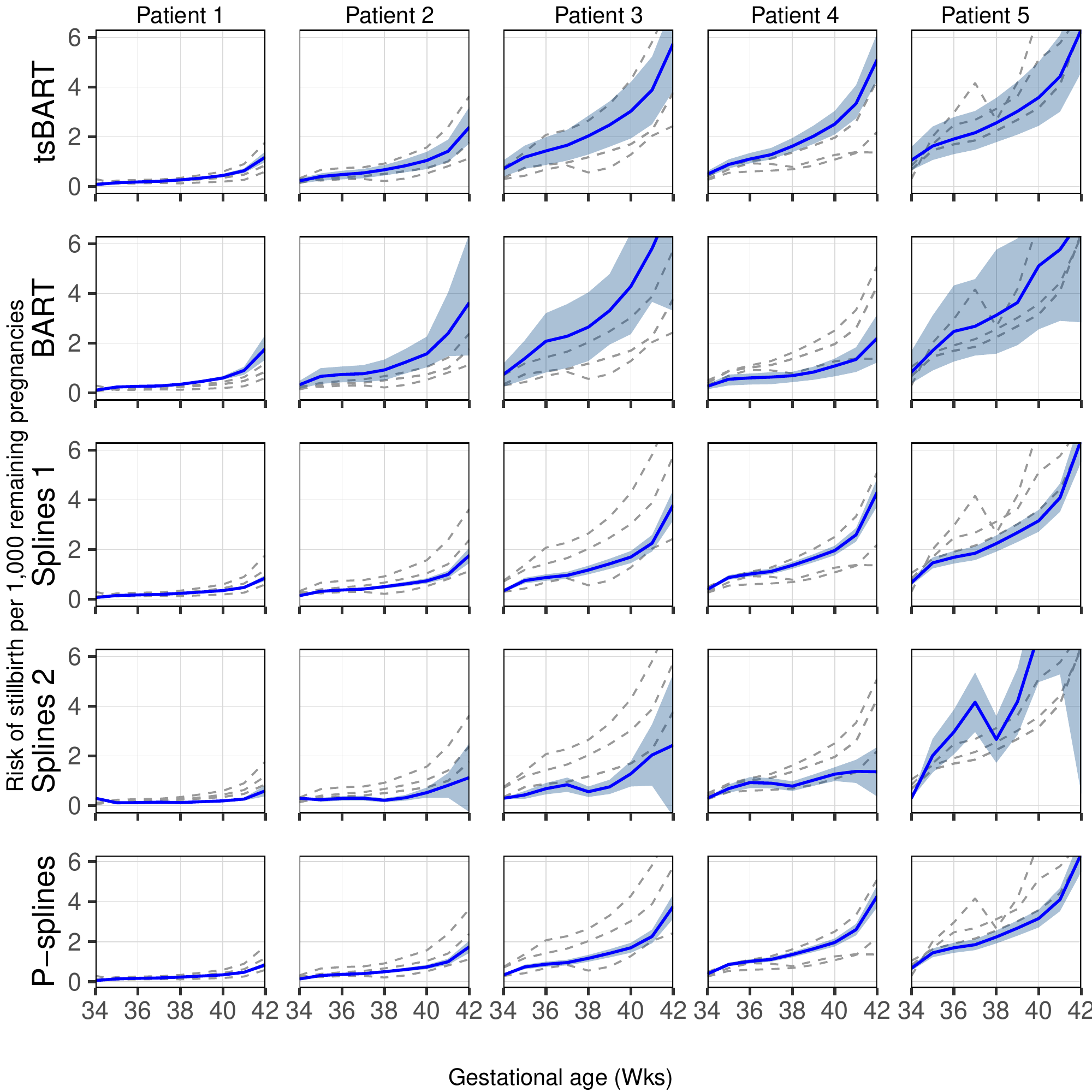}
		\caption{Estimated stillbirth risk curves for five hypothetical patients with different combinations of maternal-fetal covariates, using full case-control sample. Each row is a method, and each column is a hypothetical patient.  In each row, the posterior mean and credible interval are highlighted (dark line and shading), while the other methods' posterior means are in dashed lines for comparison.  Patient 1 is a low-risk patient (young, primiparous, no medical history, normal weight gain and birth weight).  Patient 2 introduces hypertension; Patient 3 introduces both diabetes and hypertension.  Patient 4 is multiparous with very low birth weight, and Patient 5 has a combination of risk factors (older, diabetes, hypertension, medical history, induced labor). \methAC{} gives a smooth fit with sensible credible intervals, consistent with clinical intuition about the way stillbirth risk evolves with gestational age.}
		\label{fig:panel}
	\end{figure}

We next turn to the question of how obstetricians might use the results of tsBART to understand stillbirth risk and communicate that risk to their patients.  To do so, we construct a set of hypothetical ``test'' patients representing various configurations of maternal-fetal characteristics:
\begin{itemize}
\item Patient 1 is a young, primiparous, white patient in her early 20's, with no medical history, normal weight gain, and normal birth weight for a female infant.
\item Patient 2 is otherwise similar to Patient 1, but has hypertension.
\item Patient 3 is otherwise similar to Patient 1, but has both hypertension and diabetes.
\item Patient 4 is also a young white patient in her early 20's, but is multiparous, with birth weight less than the 10th quantile.
\item Patient 5 is a white patient in her early 40's, with diabetes, hypertension, and other risk factors present; her labor is induced, and her infant is male. 
\end{itemize}

To maintain computational tractability, we again select a case-control sample of the overall data set.  We include all stillbirths in the case-control sample.  Then for each gestational age, we sample 2\% of the live births at that age.  As a result, stillbirths are 50 times more prevalent in our sample than they are in the full data set, both overall and at each gestational age.  This approach yields a dataset that is still reasonably large, with 91,078 pregnancies: all 7,175 stillbirth cases and 83,903 live-birth controls.  While we would prefer to fit the model to all 4.55 million data points, we are not yet able to do so, owing to computational constraints.  Scalable Bayesian ensemble methods are an active area of research, and we are currently drawing on this work to develop methods for scaling \methA{} to use the entire dataset.

We use this large case-control sample to fit all methods from Section \ref{sec:simstudy}.  We use the results to produce estimates of the stillbirth hazard function for each of our hypothetical test patients.  We then rescale the estimated hazard functions to account for the 50-fold down-sampling of live births in our case-control sample, and we express the resulting hazard functions as a stillbirth rate per 1,000 live births.  

The results are shown in Figure \ref{fig:panel}.  Each column represents one test patient, while each row shows a particular method.  In each panel, we show the estimated conditional probability of stillbirth risk at gestational age $t$, given survival through time $t-1$, along with 95\% uncertainty intervals.  Estimated probabilities for all other methods are also visible in grey within each panel, for easier comparison across panels.  For \methA{} and BART, the estimates are posterior-mean predicted probabilities and (Bayesian) credible intervals; for spline methods, the estimates are predicted probabilities and (frequentist) prediction intervals. 

These plots have several features of interest (we focus on the tsBART results in the top row).  First, there is considerable heterogeneity in the estimated stillbirth risk curves: in their shape, level, and degree of interaction between maternal-fetal covariates and gestational age.  Patient 1, for example, has a lower overall risk with a relatively small increase in risk at very late gestational ages (41-42).  Patients 2-4 have slightly higher overall risk at earlier gestational ages, but more much pronounced ``spikes'' in risk at late gestational ages, when the inherent stillbirth risk at an advanced stage of pregnancy is exacerbated by these patients' covariates (hypertension, diabetes + hypertension, and low fetal weight, respectively).  Patient 5, on the other hand, has a higher overall risk at all gestational ages, but a much more linear risk trajectory across gestational age compared with Patients 1-4, without the pronounced spike.

This striking heterogeneity across the patients illustrates the shortcomings of collapsing patients into two risk groups, as in \citet{mandujano2013}.  Our method, in contrast, can produce individualized estimates of risk for any patient, across all gestational ages.

We note that the estimates from the BART model are generally similar in shape to the tsBART estimates, but lack smoothness over gestational age.  This results in increased variance and poorer overall out-of-sample performance, as evident from Table \ref{tbl:binll}.  It also invites clinicians and patients to over-interpret small wiggles in the risk curves that are a result of estimation noise, rather than clinically meaningful differences.  The spline models, meanwhile, tend to result in estimates that are either over-smooth (Splines 1, P-splines) or undersmoothed and erratic (Splines 2).  We attribute this to the fact that Splines 1 and P-splines are underparametrized: they fail to include clinically meaningful interactions (e.g.~between hypertension and diabetes).  This results in higher bias, poorer estimation performance, infeasibly narrow confidence intervals that, in light of our simulation studies (Section \ref{sec:simstudy}), are likely to be anti-conservative.   Splines 2, meanwhile, is likely overparametrized: it allows for the possibility of all pairwise interactions between maternal-fetal covariates and gestational age, needlessly inflating variance for the sake of finding a small handful of clinically important interactions.  This suggests that the spline models, in order to yield good performance for stillbirth prediction, would require more nuanced model selection and attention to functional form, since including more flexible interactions was not a fruitful approach.

TsBART, in contrast, produces the best out-of-sample performance, smooth estimates, and wider, more clinically sensible error bars.  It also finds the important interactions out of the box, without the need to specify them by hand or conduct a specification search for the right form of the model.  In addition, the posterior credible  intervals from tsBART are noticeably wider for patients with unusual combinations of characteristics---an intuitive result which reflects a higher degree of uncertainty about rarer, more medically complex cases.

\section{Discussion} 

Our \methA{} model is a novel extension of BART which allows for targeted smoothing over a selected covariate.  \methAC{} enjoys the same advantages as BART: excellent predictive performance, easily tunable hyperparameters, and avoiding specification of interactions.  Hyperparameters are set efficiently via data-driven approaches using recommendations from \cite{chipman2010} and our suggested method for tuning the length-scale of the covariance function.  \methAC{} provides regularization in the form of constraining trees to be shallow learners in the prior, which is a well studied and highly successful approach to regularization in regression.


The kind of stillbirth risk analysis made possible by \methA{} represents a substantial advancement on previous work in obstetrics \citep{mandujano2013}, in terms of capturing heterogeneity of risk curves by patient and quantifying levels of certainty around each risk curve.  Further investigation into nuanced approaches for stillbirth risk modeling is warranted; maternal-fetal covariates such as age, weight gain, and birth weight may play a role in risk of stillbirth, and may interact with other covariates in complex ways.  Our fully Bayesian approach naturally allows the model to capture rich and complex interactions and quantify uncertainty about stillbirth risk, which appropriately varies by patient.

We recognize the potential limitation of confounding between the decision to induce labor, risk of stillbirth, and maternal-fetal covariates.  We currently consider the decision to induce to be a proxy for other maternal-fetal covariates which may increase stillbirth risk but are not included in the model; future work may include modeling this covariate in a causal framework.  A second limitation is the inability to link birth records to the same mother, potentially violating the independence assumption (the de-identified nature of the data prevents this linking).  However, because our data set spans only two years, it is unlikely that a large fraction of the overall births are multiple births to the same mother.  Moreover, the concern about non-independence is mitigated because we have included many of the known risk factors for stillbirth in our model.  While it is not plausible that two stillbirth events for the same mother are marginally independent, it is much more plausible that they are conditionally independent, or nearly so, given these risk factors.

Future areas of methodological work may include extension of \methA{} to a causal inference framework for observational data, as well as extension to other priors with other types of structure.  \methAC{} may be adapted in the accelerated framework of \cite{he2018} to speed computation time. It would also be interesting to explore more nuanced characterizations of partial dependence of stillbirth risk on individual covariates.  For example, plots of individual conditional expectation (ICE) may be used to assess partial relationships between response and specific covariates, using the techniques described in \citet{goldstein2015}.   ICE plots go beyond the simple partial dependence plot, by showing the functional relationship between response and feature at the level of individual observations (rather than averaging across the sample).  This could potentially give insight into the extent of potential heterogeneity in the conditional expectation function.  ICE plots can be created using the ICEbox R library \citep{goldstein2015}.

\appendix
\section{Appendix} \label{sec:appendix} 


\subsection{Review of the Bayesian backfitting MCMC} \label{appendix1}

The original BART model is typically fit using an algorithm called Bayesian backfitting \citep{hastie2000,chipman2010}.  We review this algorithm, then describe the modifications necessary to fit the \methN{} model.

Bayesian backfitting involves sampling each tree and its parameters one a time, given the partial residuals from all other $m-1$ trees.  One iteration of the sampler consists of looping through the $m$ trees, sampling each tree $T_j$ via a Metropolis step, and then sampling its associated leaf parameters $M_j$, conditional on $\sigma^2$ and the remaining trees and leaf parameters.  After a pass through all $m$ trees, $\sigma^2$ is updated in a Gibbs step.

To sample $\left\{T_j,M_j \right\}$ conditioned on the other trees and leaf parameters $\left\{T_{-j}, M_{-j} \right\}$, define the partial residual as
\begin{align}
	\label{eqn:resids}
	R_{ij} = y_i - \sum_{k=1, k \neq j}^{m} g(x_i;T_k,M_k) \, .
\end{align}
Using $R_j$ as the working response vector, at step $s$ of the MCMC one samples $T_j^{(s)}$ by proposing one of four local changes to $T_j^{(s-1)}$, marginalizing analytically over $M_j$.  The local change is selected randomly from the following candidates: 
\begin{itemize}
	\item \textbf{grow} randomly selects a terminal node and splits it into two child nodes
	\item \textbf{prune} randomly selects an internal node with two children and no grandchildren, and prunes the children, making the selected node a leaf
	\item \textbf{change} randomly selects an internal node and draws a new splitting rule
	\item \textbf{swap} randomly selects a parent-child pair of internal nodes and swaps their decision rules
\end{itemize}
The \textbf{change} and \textbf{swap} moves are computationally expensive; in practice, BART is often implemented with only \textbf{prune} and \textbf{grow} proposals \citep{pratola2014}.  Once the move in tree space is either accepted or rejected, $M_j$ is sampled from its Gaussian full conditional, given  $T_j$ and $\sigma^2$.  

\subsection{Fitting the \methA{} model with Bayesian backfitting} \label{appendix2}

Our approach to fitting \methA{} retains the form of the Bayesian backfitting MCMC algorithm, as detailed by \cite{chipman2010}.  The primary modification is that all conjugate updates are modified to their multivariate forms.  We assume an i.i.d.~error structure, although this is easily modified; and we also use a redundant multiplicative parameterization of the scale parameter, to facilitate faster MCMC mixing \citep{gelman2006, hahn2017}.   Thus our model is
\begin{align*}
	y_i &= \alpha(t_i) + \eta f(t_i,x_i) + \epsilon_{i}, \quad \epsilon_{i}(t) \stackrel{iid}{\sim} \text{N}(0,\sigma^2) \\
	f(t_i,x_i) &= \sum_{j=1}^{m}g(t_i, x_i; T_j,M_j) \; , \quad M_j = \{\mu_{1j}(t), \ldots, \mu_{b_jj}(t) \} \\
	\mu_{hj}(t) &\sim \mbox{GP}(0, C(t, t')) \\
	\eta &\sim \text{N}(\tau_0,\gamma^2) \\
	\gamma^2 &\sim \text{IG}\left(\frac{1}{2}, \frac{1}{2}\right) \\
	\sigma^2 &\sim \nu \lambda / \chi^2_\nu.
	\end{align*}
Recall that $\mu_{hj}(t)$ is the function at terminal node $l$ of tree $j$.  As described previously, this function has a Gaussian process prior with squared exponential covariance function with length scale $l$.  Because we have already introduced $\eta$ as a leading multiplicative scale parameter, we set the variance parameter of the covariance function to be $1/m$, and calibrate the prior half-Cauchy median $\tau_0$ to the marginal standard deviation of $y$.


We use the same prior for over trees  $T_j$ as in \citet{chipman2010} and \citet{hahn2017}, and so we omit many details here and refer the interested reader there.  Specifically, these papers parametrize tree depth in terms of the pair $(\alpha, \beta)$; we set $(\alpha=.95, \beta=2)$, which puts high probability on trees of depth 2 and 3, and minimizes probability on trees with depth 1 or greater than 4.  For $\sigma^2$, we follow Chipman et al.'s recommendation for a rough over-estimation of $\hat{\sigma}$.  We choose $\nu=3$ and $q = 0.90$, and estimate $\hat{\sigma}$ by regressing $y$ onto $x$ (including the index variable as a covariate), then choose $\lambda$ s.t. the $q$th quantile of the prior is located at $\hat{\sigma}$, i.e. $P(\sigma \le \hat{\sigma}) = q$.

The posterior conditional distributions are as follows.  For simplicity of notation, we assume times $t$ are on a common discrete grid, where T is again the range of $t$ values in the data set (although this is not a requirement of the method). We update $\sigma^2$ as
\begin{align*}
	\sigma^2 \mid \ldot \sim
	\frac{\nu\lambda + \text{RSS}}{\chi^2_{\nu + N + 1}} \quad \text{where } \text{RSS} = \sum_{i, t}\left(y_i(t) -  \eta f(t_i,x_i)\right)^2.
\end{align*}
where $N$ is the count of observations across all time points, $N = \sum_{i=1}^n N_i$ where $N_i$ is the number of time points for observation $i$, and $\chi^2_{\nu + N + 1}$ is a draw from a chi-squared random variable.

The update for each $\mu_h = \left[\mu_h^{(1)}, \hdots, \mu_h^{(T)} \right]$ is
\begin{align*}
	\mu_h \mid \ldot \sim
	\text{N}\left( \tilde{m}, \tilde{\Sigma} \right) \hspace{0.5cm} \text{where }
	\tilde{\Sigma} = \left(\Lambda + K \right)^{-1} \text{ and } \tilde{m} = \tilde{\Sigma} \left(\Lambda \bar{y}_l + K \mu_0 \right)
\end{align*}
where $\Lambda = N_l^{-1}$ is the inverse of the diagonal matrix of sample sizes for each time point for observations in leaf $l$, $K = \Sigma_0^{-1}$, and $\bar{y}_l$ is the vector of sample means for observations in leaf $l$ at each time point.

The update for $\eta$ is Gaussian,
\begin{align*}
	\mu_h \mid \ldot &\sim
	\text{N}(\tilde{m},\tilde{v}^2) \hspace{0.5cm} \text{where }\\ 
	\tilde{v}^2 &= \left( \frac{\tau_0}{\gamma^2}  + \frac{1}{\sigma^2}\sum_{i,t} f(t_i,x_i)^2 \right)^{-1}\\
	\tilde{m} &= \tilde{v}^2\left( \frac{\tau_0}{\gamma^2}  + \frac{1}{\sigma^2} \sum_{i,t} y_i f(t_i,x_i) \right).  
\end{align*}

Finally, the update for $\gamma^2$ is
\begin{align*}
	\gamma^2 \mid \ldot \sim
	 \text{IG}\left(1, \frac{\eta^2+1}{2} \right).
\end{align*}

For updating the trees $T_j$, the marginal likelihood is the corresponding multivariate extension to the marginal likelihood in regular BART.  We again let $R_{ij}$ represent the partial residuals as defined in Equation $\ref{eqn:resids}$, and let $R_l$ denote the vector containing residuals for data points in leaf $l$.  We then obtain the marginal likelihood for the $b$ terminal nodes as
\begin{align*}
	p(R_h \mid T_j, M_j, \sigma^2) = 
	\int_{\mu_h} \prod_{l \in 1:b} 
	\text{N}\left(R_h \mid W_l \mu_h, \sigma^2 I \right)
	\cdot
	\text{N}\left(\mu_h \mid \mu_0, \Sigma_0 \right)
	\partial \mu_h
\end{align*}
where $W_l$ is a $(t_\text{max} \times n)$ matrix where elements indicate times at which each $y_{i}$ is observed.  This Gaussian integral is easily computed in closed form.

\subsection{Additional detail on hyperparameter tuning for length-scale} \label{appendix3}
	
Here we provide additional detail regarding tuning the expected number of crossings $\text{E}\left[N_T(0) \right]$ for calculating the covariance's length-scale parameter.  We select the optimal $\text{E}\left[N_T(0) \right]$ by beginning with a grid of candidate values $e_c \in \left\{e_1,\hdots,e_C \right\}$.  For each candidate $e_c$, we fit the \methN{} model and calculate WAIC \citep{watanabe2013}, yielding a grid of WAIC values $\Omega = \left\{\omega_1, \hdots, \omega_C \right\}$.
	
\begin{figure}
		\centering
			\includegraphics[scale=.5]{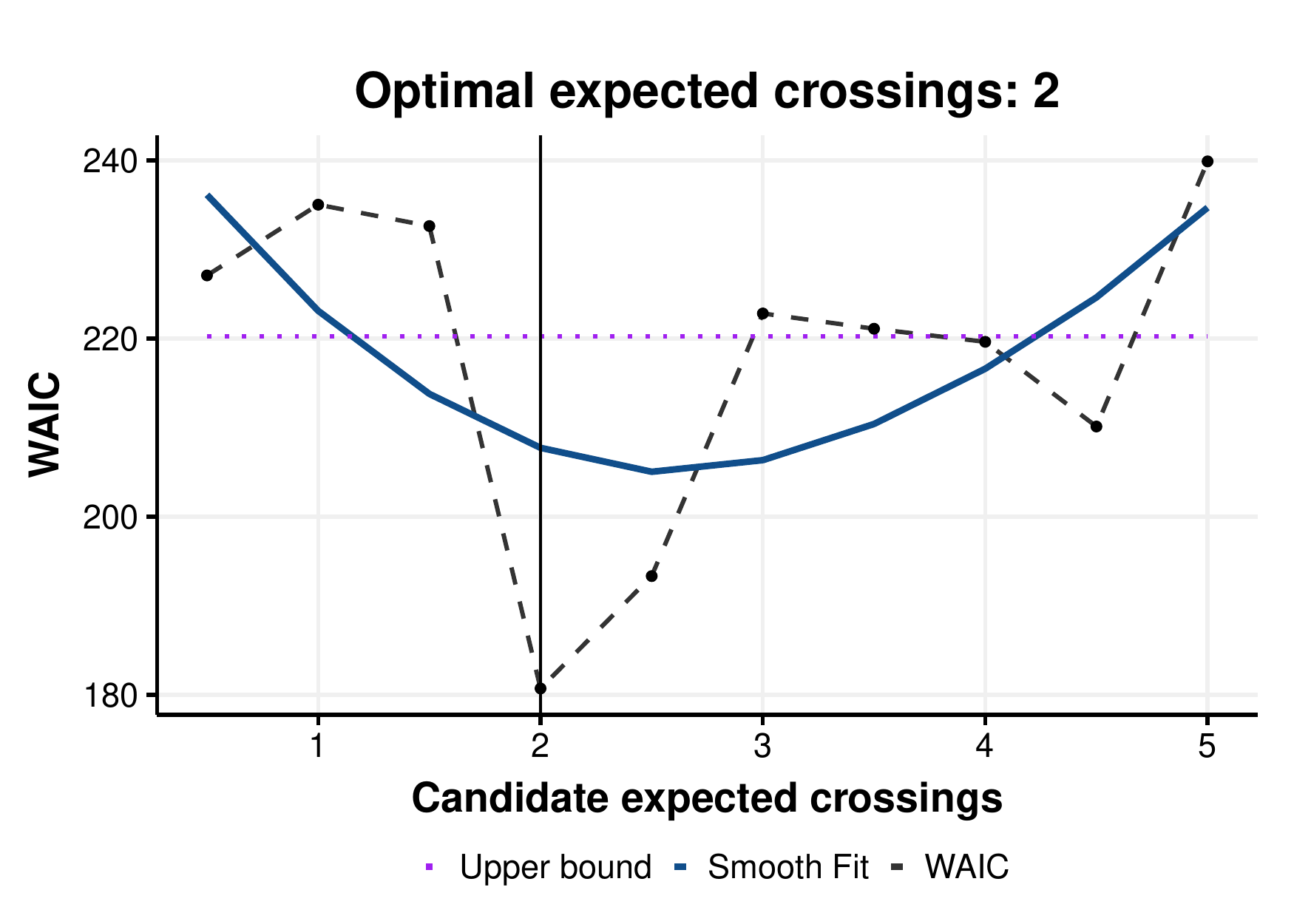}
		\caption{Example of tuning the expected number of crossings.  The jagged dashed line illustrates the Monte Carlo variation present in WAIC estimates.  The solid line shows the spline fit.  The horizontal dotted line shows the minimum WAIC value plus one standard deviation from the spline fit.  The solid vertical line gives the minimum candidate expected number of crossings value where there is a WAIC value less than one plus the standard deviation.}
		\label{fig:ec}
	\end{figure}

The WAIC values contain Monte Carlos variation; to overcome this, we fit a cubic spline model to $\Omega$.  Let $\zeta$ be the standard deviation of the residuals from this model fit.  We select the smallest number of expected crossings $e_c$ where the corresponding $\omega_c$ is within $\zeta$ of $\min(\Omega)$.  This approach encourages smoothing while maintaining performance.  Figure \ref{fig:ec} gives a visualization of this tuning.  Other methods such as cross-validation could easily be used for tuning the expected number of crossings; we find this data-driven approach to be efficient while still yielding good results.

\subsection{Simulation Details} \label{appendix4}

Here we provide more detail for the second simulation described in Section \ref{sec:simstudy}.  We simulate data as follows.  Let $t$ be a grid of times on the unit interval, spaced in increments of 0.1.  We generated $n=1000$ ten-dimensional covariates $x_i = \left\{x_{i1}, \hdots, x_{i10} \right\}$ where $x_{ij} \stackrel{iid}{\sim} U(0,1)$.  The first five covariates in each $x_i$ impact the response; the rest are noise.

We generate data using a weighted combination of two risk functions, where $f_1(t) = 0.075t$ is the baseline risk function, and $f_2(t)=0.75 \cdot \max(0.75, t)^{\rho}$ is a second risk function which controls a large `kick` at $t=0.75$.  We let $\rho = 1 + \log(0.1) / \log(0.75)$, so that the $f_2(t)$ risk at $t=1$ is five times the baseline risk.

The weights $w(x_i)$ for combining $f_1(t)$ and $f_2(t)$ are dependent on the covariates $x_i$.  We generate data for three scenarios, letting weights $w(x_i)$ depend on covariates in either linearly, linearly with interactions, or nonlinearly with interactions.  These scenarios represent increasing degrees of difficulty in learning the underlying function.

\begin{itemize}
	\item Linear: 
	\begin{align*}
		w(x_i) = \text{sigmoid}\left[5\left(x_{i1} - x_{i2} + x_{i3} - x_{i4} \right) \right]
	\end{align*}
	\item Linear with interaction: 
	\begin{align*}
		w(x_i) = \text{sigmoid}[
					&5\left(x_{i1} - x_{i2} \right) + 
				 	5\left(x_{i1}-0.5 \right)\left(x_{i2}-.5 \right) +\\
					&5\left(x_{i3}-x_{i4} \right) +
					5\left(x_{i3}-.5 \right)\left(x_{i4}-.5 \right)]
	\end{align*}
	\item Nonlinear with interaction:
	\begin{align*}
		w(x_i) = \text{sigmoid}\left[
					5\left( \max\left(x_{i1},x_{i2}\right) \right) - 
					5\left( \max\left(x_{i3},x_{i4} \right)\right)
				 \right]
	\end{align*}
\end{itemize}

We then generate the simulated hazard function data according to $h(t)$, rescale responses so that the overall survival probability is roughly 0.5, and simulate event times for each observation.
\begin{align*}
	h(t) = 0.25x_{i5} + w(x_i)f_2(t) + (1-w(x_i))f_1(t)
\end{align*}

\subsection{Stillbirth Results Using Random Forest} \label{appendix5}
\begin{figure}
		\centering
			\includegraphics[width=0.9\textwidth]{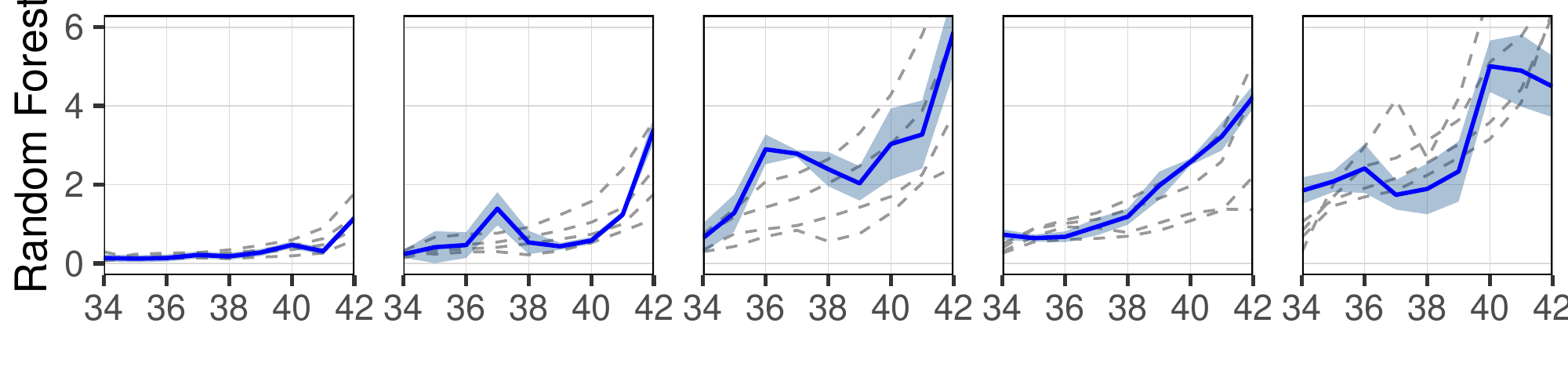}
		\caption{Estimated stillbirth risk curves from the Random Forest model, using the same five hypothetical patients and full case-control sample as Figure 5. The posterior mean and credible intervals are highlighted (dark line and shading), while the other methods' posterior means are in dashed lines for comparison.  Each column is a hypothetical patient.  Patient 1 is a low-risk patient (young, primiparous, no medical history, normal weight gain and birth weight).  Patient 2 introduces hypertension; Patient 3 introduces both diabetes and hypertension.  Patient 4 is multiparous with very low birth weight, and Patient 5 has a combination of risk factors (older, diabetes, hypertension, medical history, induced labor). Random Forest gives curves generally similar in shape to BART, and does not induce smoothness.
				}
		\label{fig:fig7}
	\end{figure}
	
Here we illustrate the Random Forest fit for the stillbirth dataset.  We do not include Random Forest in the set of models for stillbirth analysis; while \citet{wager2014} provide variance estimation for Random Forest, \citet{chipman2010} demonstrated that BART tends to outperform Random Forest.  In addition, Random Forest does not induce smoothness, as we see in Figure \ref{fig:fig7}.

\section{Appendix}
The R package \textbf{tsbart} implements the BART with Targeted Smoothing method.  It is available at \url{https://github.com/jestarling/tsbart}.


\bibliographystyle{abbrvnat}
\bibliography{tsbart-paper}

\end{document}